%% file: main.tex
\documentclass[sigconf, nonacm]{acmart}

\AtBeginDocument{%
  }

\usepackage{acronym}
\usepackage{amsmath}
\usepackage[english]{babel}
\usepackage[shortlabels,inline]{enumitem}
\usepackage{graphicx}
\usepackage{hyperref}
\usepackage{cleveref}
\usepackage[utf8]{inputenc}
\usepackage{lipsum}
\usepackage{siunitx}
\usepackage{xcolor}
\usepackage[T1]{fontenc}
\usepackage{threeparttable}
\usepackage{csquotes}
\usepackage{bbm}
\usepackage{mathtools}
\usepackage{microtype}
\input{acronyms}

\begin{document}
\title{Matched Filter-Based Molecule Source Localization in Advection-Diffusion-Driven Pipe Networks with Known Topology}
\author{Timo Jakumeit\textsuperscript{1,*}\!, Bastian Heinlein\textsuperscript{1,2,*}\!, Vukašin Spasojević\textsuperscript{1}\!, Vahid Jamali\textsuperscript{2}, Robert Schober\textsuperscript{1}\!, and Maximilian Sch\"afer\textsuperscript{1}}
\affiliation{%
  \institution{\textsuperscript{1} Friedrich-Alexander-Universit\"at Erlangen-N\"urnberg (FAU), Erlangen, Germany} \city{}\country{}
}
\affiliation{%
  \institution{\textsuperscript{2} Technical University of Darmstadt, Darmstadt, Germany} \city{}\country{}
}
\thanks{\textsuperscript{*}Both authors contributed equally to this work.\\
This work was funded in part by the German Federal Ministry of Research, Technology and Space (BMFTR) through Project Internet of Bio-Nano-Things (IoBNT) -- grant numbers 16KIS1987 and 16KIS1992, in part by the German Research Foundation (Deutsche Forschungsgemeinschaft, DFG) under GRK 2950 -- ProjectID 509922606 and under grant number SCHA 2350/2-1, in part by the European Union’s Horizon Europe -- HORIZON-EIC-2024-PATHFINDEROPEN-01 under grant agreement Project N. 101185661, and in part by the Horizon Europe Marie Skodowska Curie Actions (MSCA)-UNITE under Project 101129618.}

\renewcommand{\shortauthors}{Jakumeit and Heinlein, et al.}

\newcommand{\defeq}{\vcentcolon=}

\begin{abstract}
Synthetic \ac{MC} has emerged as a powerful framework for modeling, analyzing, and designing communication systems where information is encoded into properties of molecules. 
Among the envisioned applications of \ac{MC} is the localization of molecule sources in \acp{PN} like the human \ac{CVS}, \acp{SN}, and industrial plants.
While existing algorithms mostly focus on simplified scenarios, in this paper, we propose the first framework for source localization in complex \acp{PN} with known topology, by leveraging the \textit{\ac{MIGHT}} model as a closed-form representation for advection-diffusion-driven \ac{MC} in \acp{PN}. 
We propose a \ac{MF}-based approach to identify molecule sources under realistic conditions such as unknown release times, random numbers of released molecules, sensor noise, and limited sensor sampling rate.
We apply the algorithm to localize a source of viral markers in a real-world \ac{SN} and show that the proposed scheme outperforms randomly guessing sources even at low \acp{SNR} at the sensor and achieves error-free localization under favorable conditions, i.e., high \acp{SNR} and sampling rates. Furthermore, by identifying clusters of frequently confused sources, reliable cluster-level localization is possible at substantially lower \acp{SNR} and sampling rates.
\end{abstract}
\acresetall

\maketitle

\section{Introduction}\label{sec:Introduction}
\input{sections/Introduction}

\section{System and Channel Model}\label{sec:System_Model}
\input{sections/system_model}

\section{Molecule Source Localization}\label{sec:Localization_Algorithms}
\input{sections/localization_algorithms_new}

\section{Results}\label{sec:Results}
\input{sections/results}

\section{Conclusion}\label{sec:Conclusion}
\input{sections/conclusion}

\bibliographystyle{unsrtnat}
\bibliography{bibliography}

\end{document}

%% file: acronyms.tex
\acrodef{0-D}[0-D]{zero-dimensional}
\acrodef{1-D}[1-D]{one-dimensional}
\acrodef{2-D}[2-D]{two-dimensional}
\acrodef{3-D}[3-D]{three-dimensional}
\acrodef{AWGN}[AWGN]{additive white Gaussian noise}
\acrodef{ACF}[ACF]{autocorrelation function}
\acrodef{ARI}[ARI]{adjusted rand index}
\acrodef{LP}[LP]{low-pass}
\acrodef{MC}[MC]{molecular communication}
\acrodef{MF}[MF]{matched filter}
\acrodef{CM}[CM]{confusion matrix}
\acrodef{CSM}[CSM]{cosine similarity matrix}
\acrodef{SN}[SN]{sewage network}
\acrodef{EM}[EM]{electromagnetic wave}
\acrodef{CGF}[CGF]{cumulant generating function}
\acrodef{CVS}[CVS]{cardiovascular system}
\acrodef{CIR}[CIR]{channel impulse response}
\acrodef{FPT}[FPT]{first passage time}
\acrodef{IoBNT}[IoBNT]{Internet of Bio-Nano Things}
\acrodef{IG}[IG]{inverse Gaussian}
\acrodef{ISI}[ISI]{inter-symbol interference}
\acrodef{MIGHT}[MIGHT]{mixture of inverse Gaussians for hemodynamic transport}
\acrodef{MIMO}[MIMO]{multiple-input multiple-output}
\acrodef{MISO}[MISO]{multiple-input single-output}
\acrodef{SIMO}[SIMO]{single-input multiple-output}
\acrodef{SISO}[SISO]{single-input single-output}
\acrodef{PDF}[PDF]{probability density function}
\acrodef{RV}[RV]{random variable}
\acrodef{SNR}[SNR]{signal-to-noise ratio}
\acrodef{Tx}[Tx]{transmitter}
\acrodef{Rx}[Rx]{receiver}
\acrodef{UCA}[UCA]{uniform concentration assumption}
\acrodef{VN}[VN]{vessel network}
\acrodef{PN}[PN]{pipe network}
\acrodef{wrt}[w.r.t.]{with respect to}
\acrodefplural{ACF}[ACFs]{autocorrelation functions}
\acrodefplural{IG}[IGs]{inverse Gaussians}
\acrodefplural{PDF}[PDFs]{probability density functions}
\acrodefplural{RV}[RVs]{random variables}
\acrodefplural{SNR}[SNRs]{signal-to-noise ratios}
\acrodefplural{GFP}[GFPs]{green fluorescent proteins}
\acrodefplural{CIR}[CIRs]{channel impulse responses}
\acrodefplural{VN}[VNs]{vessel networks}
\acrodefplural{PN}[PNs]{pipe networks}
\acrodefplural{Tx}[Txs]{transmitters}
\acrodefplural{Rx}[Rxs]{receivers}
\acrodefplural{SN}[SNs]{sewage networks}
\acrodefplural{MFs}[MFs]{matched filters}

%% file: sections/introduction.tex
\acresetall
In the past years, synthetic \ac{MC} has emerged as a powerful framework for modeling, analyzing, and designing information exchange in systems where molecules act as information carriers, and numerous innovative applications within the human body~\cite{Mosayebi2019}, between plants~\cite{Unluturk2017}, or in industrial environments have been envisioned~\cite{Felicetti2016, Farsad2016, Bi2021}.

One application that has attracted sustained interest is anomaly localization in \ac{MC} systems~\cite{Etemadi2023}. Especially \textit{molecule source localization}, i.e., the estimation of the location of a \ac{Tx} emitting signaling molecules within an \ac{MC} system, has proven particularly relevant for various application scenarios. 
In biomedical applications, source localization shall enable the localization of diseased tissue within the human body or more precisely, the \ac{CVS}, with the potential to identify pathologies much earlier than classical medical imaging techniques~\cite{Mosayebi2019}. 
In epidemiological surveillance, source localization based on molecular signals in \acp{SN} can support the early detection and spatial attribution of infections, offering a cost-effective alternative to population-wide testing~\cite{Deng2022}. 
In industrial systems, molecule source localization could provide a complementary approach for identifying leaks or contaminant releases in \acp{PN} by exploiting concentration measurements of specific chemical species.

The aforementioned application domains span different size scales (see Fig.~\ref{fig:system_model}a)) but share a common structural characteristic: they can be represented as \acp{PN}, i.e., networks of interconnected conduits through which molecules are transported by flowing liquids or gases.
In the \ac{CVS}, blood vessels form \acp{PN} carrying biomarkers, nutrients, and metabolic waste.
In \acp{SN}, pipes transport wastewater and potential infection markers.
In industrial gas or fluid networks, pipes convey process media, while released chemicals, pollutants, or tracers propagate through the \ac{PN} and are detectable downstream.
\acp{PN} thus represent a particularly important class of environments for molecule source localization. 
However, the structural features of \acp{PN} induce complex molecule transport dynamics.
Therefore, accurate molecule source localization in \acp{PN} requires mathematical models that explicitly account for advection, diffusion, and the underlying network topology~\cite{Jakumeit2025b}.

Most of the existing literature on molecule source localization in \ac{MC} is focused on unbounded diffusion environments, and only a limited number of studies have considered source localization in single pipes or individual pipe branchings, where molecule propagation is governed by advection and diffusion.
In~\cite{Turan2018}, a \ac{1-D} analytical source localization approach for a single pipe is presented, where the source (i.e., \ac{Tx}) position is inferred via \ac{Tx}-\ac{Rx} distance estimation. 
In~\cite{Khaloopour2021}, a \ac{1-D} analytical framework for cooperative abnormality detection and localization is proposed, in which mobile sensors propagate in a single pipe and release signaling molecules upon sensing an abnormality. 
In~\cite{Schottlender2025}, a learning-based distance estimation approach for branched \ac{MC} systems is proposed, where multiple \acp{Tx} emit molecules that propagate by advection-only through a simple Y-shaped pipe topology towards the \ac{Rx}. Recurrent neural networks are used to infer \ac{Tx}-\ac{Rx} distances from the received signal. 
In~\cite{Pal2025}, a learning-based framework for infection source localization in the \ac{CVS} is proposed. Using the BloodVoyagerS simulator, biomarkers released by static infection sources propagate through a \ac{PN} of the major arteries via advection and subject to chemical degradation. Localization is formulated as a multi-class classification problem based on received biomarker signals using simulation data and stacked ensemble learning.

In summary, existing works on molecule source localization in \ac{MC} predominantly consider highly simplified channel topologies and transport dynamics. 
Most approaches are limited to diffusion-only or advection-only models, with only a few studies accounting for their combined effects.
Moreover, no existing works provide analytically tractable and physically interpretable localization frameworks that are applicable to \acp{PN} across different scales and different application scenarios. 
To overcome the drawbacks of existing approaches, we propose the first molecule source localization framework for advective-diffusive \ac{MC} in \acp{PN}. 
Our approach builds on the \textit{\ac{MIGHT}} model~\cite{Jakumeit2026}, which provides a closed-form description of molecule transport in \acp{PN} and is extended here to explicitly account for key system uncertainties, including unknown release times and quantities, sensor noise, and different sampling rates at the \ac{Rx}. 
Based on this model, we develop a \ac{MF}-based localization algorithm and demonstrate that clustering frequently confused \acp{Tx} can significantly enhance localization performance on a cluster-level. 
The effectiveness of the proposed framework is illustrated by applying it for the localization of a source of viral markers in a real-world \ac{SN} in the Zolkiewka Commune in Poland, highlighting its practical relevance and applicability.

The main contributions of this work are as follows:
\begin{itemize}
    \itemsep0em
    \item We extend the deterministic \ac{MIGHT} channel model from \cite{Jakumeit2026} to account for log-normal-distributed random release quantities at the \acp{Tx} and sensor noise at the \ac{Rx}.  
    \item We propose an \ac{MF}-based algorithm for molecule source localization in advection-diffusion-driven \acp{PN} of arbitrary sizes and structural complexities.
    \item We introduce methods for the clustering of commonly confused \acp{Tx} to improve cluster-level localization accuracy.
    \item We illustrate the accuracy of the proposed localization framework for different sensor sampling rates by its application for source localization in a real-world \ac{SN}.
\end{itemize}

The remainder of the paper is organized as follows:
Section~\ref{sec:System_Model} introduces the system model, reviews the
\ac{MIGHT} model, and presents the \ac{Rx} noise model.
Section~\ref{sec:Localization_Algorithms} presents the \ac{MF}-based
localization approach and the \ac{Tx} clustering method.
Section~\ref{sec:Results} discusses numerical results, and
Section~\ref{sec:Conclusion} concludes the paper.

%% file: sections/system_model.tex
In the following, we first define \acp{PN}, then describe the assumptions on the molecule sources, i.e., \acp{Tx}, including the randomness of the number of released molecules, and the advective-diffusive transport.
Subsequently, we review the \ac{MIGHT} channel model.
Finally, we discuss the \ac{Rx} model and the associated sensor noise.

\begin{figure*}
    \centering
    \includegraphics[width=\linewidth]{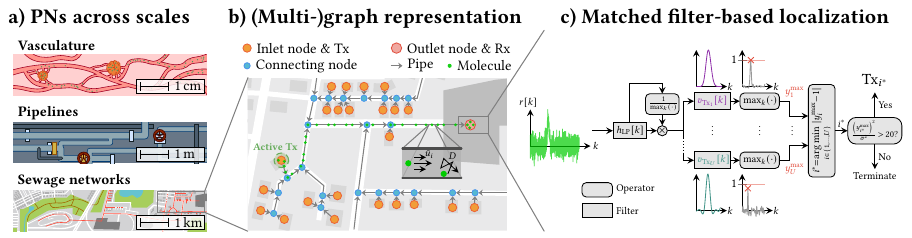}\vspace*{-3mm}
    \caption{System model. a)~Advection-diffusion-driven \acp{PN} across different scales. b)~(Multi-)graph representation of a \ac{PN}. c)~\ac{MF} bank-based molecule source localization algorithm, exploiting the \ac{MIGHT} model for \ac{MF} design.}
    \Description[]{}
    \label{fig:system_model}
\end{figure*}

\subsection{Pipe Network Definition}
 
To model molecule transport in \acp{PN}, we adopt the definition from \cite{Jakumeit2026} and approximate the \ac{PN} topology using three segment types\footnote{Please refer to Fig.~2a) in~\cite{Jakumeit2026} for a visual illustration of the segment types.}:
\begin{enumerate}
    \itemsep0em
    \item \textit{Pipe:} A pipe $p_i$ is a cylindrical conduit transporting fluid from its inlet to its outlet, with length $l_i$ and radius $r_i$. Pipes connect to other pipes, bifurcations, or junctions. The \ac{PN} contains $E$ pipes.
    \item \textit{Bifurcation:} A bifurcation $b_m$ is a connection where one or more inflow pipe(s) split(s) into multiple outflow pipes. The set of its outflow pipes is denoted by $\mathcal{O}(b_m)$. A \ac{PN} contains $B$ bifurcations.
    \item \textit{Junction:} A junction is a connection where multiple inflow pipes merge into one outflow pipe.
\end{enumerate}
Bifurcations, junctions, inlet(s), outlet(s), and any connection point are modeled as nodes. We differentiate between three node types:
\begin{enumerate}
\itemsep0em
    \item \textit{Inlet node:} Inlet nodes $n_{\mathrm{in},a}\in \mathcal{N}_\mathrm{in}=\{n_{\mathrm{in},1},\ldots ,n_{\mathrm{in},I}\}$ with $\,I \in \mathbb{N}$ exist at the points of the \ac{PN} where fluid flow is introduced into the \ac{PN}. Here, $\mathbb{N}$ denotes the set of natural numbers.
    \item \textit{Outlet node:} Outlet nodes $n_{\mathrm{out},b}\in \mathcal{N}_\mathrm{out}=\{n_{\mathrm{out},1},\ldots ,n_{\mathrm{out},O}\}$ with $\,O \in \mathbb{N}$ exist at the points of the \ac{PN} where fluid flow leaves the \ac{PN}.
    \item \textit{Connecting node:} All other $C\in\mathbb{N}$ points in the \ac{PN} where pipes are connected to one another are referred to as connecting nodes $n_i\in\mathcal{N}_\mathrm{con}=\{n_{1},\ldots ,n_{C}\}$. 
\end{enumerate}
Pipes are represented as directed edges between nodes, aligned with the fluid flow direction, determined in Section~\ref{ssec:advective-diffusive_molecule_transport}.
For any node type, the node at the inlet and outlet of a pipe $p_i$, i.e., the \textit{source} and \textit{destination node}, is denoted by $\mathcal{S}(p_i)$ and $\mathcal{D}(p_i)$, respectively\footnote{Please refer to Figs.~1a) and 1c) in~\cite{Jakumeit2026} for a visual illustration of the notation.}.

The representation based on nodes and directed edges allows any \ac{PN} to be described as a directed multigraph.
The set of all distinct directed paths between a given (inlet/connecting) node $n_{a}\in\mathcal{N}_\mathrm{in}\cup \mathcal{N}_\mathrm{con}$ and another (connecting/outlet) node $n_{b}\in\mathcal{N}_\mathrm{con}\cup \mathcal{N}_\mathrm{out}$ is denoted by $\mathcal{P}(n_{a}, n_{b})$.
Each path $P_k$ comprises a subset of pipes and bifurcations given by
\begin{equation}\label{eqn:PathSet}
P_k = \left\lbrace p_i \mid i \in \mathcal{E}_k \right\rbrace \cup \left\lbrace b_m \mid m \in \mathcal{B}_k \right\rbrace,
\end{equation}
where $\mathcal{E}_k \subseteq \left\lbrace 1,\ldots ,E \right\rbrace$ and $\mathcal{B}_k \subseteq \left\lbrace 1,\ldots ,B \right\rbrace$ are the index sets of the pipes and bifurcations\footnote{Potential bifurcations at $n_a$ or $n_b$ are \textit{excluded} from the set in~\eqref{eqn:PathSet}, as molecules in any path from $n_a$ and $n_b$ do not travel \textit{through} $n_a$ or $n_b$, but start at $n_a$ and end at $n_b$.} included in $P_k$. Any path must contain at least two pipes, i.e., $|\mathcal{E}_k|>1$, where $|\cdot|$ denotes the cardinality. 

\subsection{Molecule Sources}
We assume that there are $U$ \acp{Tx}. Any $\mathrm{Tx}_g\in\mathcal{N}_\mathrm{Tx}=\{\mathrm{Tx}_1,\ldots,\mathrm{Tx}_U \}$ is a \ac{0-D} point at position $z_{\mathrm{Tx}_g}\in[0,l_i]$ in pipe $p_i$. We denote the source node of pipe $p_i$ containing $\mathrm{Tx}_g$ as $\mathcal{S}_\mathrm{Tx}(\mathrm{Tx}_g)$.
Moreover, we assume that each $\mathrm{Tx}_g$ can impulsively release $M_g$ molecules.
Here, the number of released molecules $M_g$ is modeled as a \ac{RV} with mean $\bar{M}_g$, realization $m_g$, and \ac{PDF} $f_{M_g}(m_g)$, chosen according to the application scenario (see Section~\ref{sec:Results}).

\subsection{Advective-Diffusive Molecule Transport}\label{ssec:advective-diffusive_molecule_transport}

In many practical \acp{PN}, including \acp{SN}, the \ac{CVS}, and gas pipelines, see Fig.~\ref{fig:system_model}a), molecule transport is governed primarily by advection, i.e., transport with the bulk fluid motion, and diffusion, see Fig.~\ref{fig:system_model}b), encompassing both molecular and turbulent dispersive effects. We summarize the advective-diffusive transport assumptions underlying the proposed model below.

Each inlet node $n_{\mathrm{in},a}\in\mathcal{N}_\mathrm{in}$ is assigned a flow rate $Q_{\mathrm{in},a}>0$, inducing flow in each pipe $p_i$ in the \ac{PN}, characterized by the flow rate $Q_i$ and the cross-sectional average velocity
$\bar{u}_i = Q_i/(\pi r_i^2)$ as determined using an equivalent electrical circuit model~\cite{Jakumeit2026}.
Additionally, molecules propagate via diffusion, characterized by the effective diffusion coefficient in pipe $p_i$~\cite[Eq.~(26)]{Aris1956}
\begin{equation}\label{eqn:ArisTaylorEffectiveDiffusionCoefficient}
\bar{D}_i = \dfrac{r_i^2 \bar{u}_i^2}{48 D} + D,
\end{equation}
where $D$ denotes the diffusion coefficient.

\subsection{MIGHT Channel Model}

In this work, we adopt \ac{MIGHT}~\cite{Jakumeit2026} to model the \ac{PN}.
\ac{MIGHT} exploits that the \acp{FPT} of molecules in advective-diffusive channels follow an \ac{IG} distribution, and yields tractable, closed-form expressions for molecule transport in \acp{PN}, making it well-suited for applications that require a mathematically well-behaved framework, such as source localization~\cite{Jakumeit2026}. Below, we briefly introduce the core equations of \ac{MIGHT}. 

\subsubsection{Path Molecule Flux}

Let $\mathrm{Tx}_g\in\mathcal{N}_\mathrm{Tx}$ be at position $z_{\mathrm{Tx}_g}$ in pipe $p_q$ and let the \ac{Rx} be at position $z_{\mathrm{Rx}}$ in pipe $p_w$.
Moreover, any path contained in $\{P_k\in \mathcal{P}(\mathcal{S}_\mathrm{Tx}(\mathrm{Tx}_g),\mathcal{D}(p_w))\vert p_q,p_w\in P_k\}$ leads from $\mathrm{Tx}_g$ to the \ac{Rx} and contains pipes $p_q$ and $p_w$.
Then, the molecule flux observed at the \ac{Rx} position $z_{\mathrm{Rx}}$ in pipe $p_w$ due to the release of $m_g=1$ molecule at $\mathrm{Tx}_g$ follows an \ac{IG} distribution and is obtained in $\SI{}{\per\second}$ as~\cite[Eqs.~(8), (14), (15), (17)]{Jakumeit2026}
\begin{align}\label{eqn:path_molecule_flux} 
    \bar{j}_k(\hspace*{-.3mm}z_\mathrm{Rx},t;z_{\mathrm{Tx}_g}\hspace*{-.3mm}) \hspace*{-.7mm}=\hspace*{-.7mm}\frac{\bar{\mu}_k(z_\mathrm{Rx};z_{\mathrm{Tx}_g})}{\sqrt{2\pi\bar{\theta}_k(\hspace*{-.3mm}z_\mathrm{Rx};z_{\mathrm{Tx}_g}\hspace*{-.3mm}) t^3}}\hspace*{-.5mm}\exp \hspace*{-.5mm}\left( \hspace*{-.5mm}-\frac{(\hspace*{-.3mm}t\hspace*{-.5mm}-\hspace*{-.5mm}\bar{\mu}_k(z_\mathrm{Rx};z_{\mathrm{Tx}_g})\hspace*{-.3mm})^2}{2\bar{\theta}_k(z_\mathrm{Rx};z_{\mathrm{Tx}_g}) t}\hspace*{-.5mm}\right)\hspace*{-.3mm},
\end{align}
with mean, variance, and scale parameter of path $P_k$ given by~\cite[Eq.~(15), (16), (17)]{Jakumeit2026}
\begin{align}
        &\bar{\mu}_k(z_\mathrm{Rx};z_{\mathrm{Tx}_g})=
        \mu_q(l_q-z_{\mathrm{Tx}_g})+\mu_w(z_\mathrm{Rx})+\sum_{i\in \mathcal{E}_k\backslash \left\lbrace q,w\right\rbrace}\mu_i(l_i),\label{eq:mom-het-mu}\\
         &\bar{\sigma}_k^2(z_\mathrm{Rx};z_{\mathrm{Tx}_g})
        = \sigma_q^2(l_q-z_{\mathrm{Tx}_g})+\sigma_w^2(z_\mathrm{Rx})+\sum_{i\in \mathcal{E}_k\backslash \left\lbrace q,w\right\rbrace}\sigma_i^2(l_i),\label{eq:mom-het-sigma}\\
        &\bar{\theta}_k(z_\mathrm{Rx};z_{\mathrm{Tx}_g})=\frac{\bar{\sigma}_k^2(z_\mathrm{Rx};z_{\mathrm{Tx}_g})}{\bar{\mu}_k(z_\mathrm{Rx};z_{\mathrm{Tx}_g})},
\end{align}
and mean and variance of pipe $p_i$ given by~\cite[Eq.~(7)]{Jakumeit2026}
\begin{align}
    &\mu_i(z_i) = \frac{z_i}{\bar{u}_i}, 
    &\sigma_i^2(z_i) = \frac{2\bar{D}_i z_i}{\bar{u}_i^3},
    \label{eqn:pipe_mean_variance_scale}
\end{align}
where $z_i\in[0,l_i]$ denotes the longitudinal position within pipe $p_i$.

\subsubsection{Channel Impulse Response}

Given the path molecule flux in~\eqref{eqn:path_molecule_flux},
the \ac{CIR} between $\mathrm{Tx}_g$ and position $z_\mathrm{Rx}$ in pipe $p_w$ in $\SI{}{\per\second}$ is obtained by summing up the weighted path fluxes of all paths between $\mathrm{Tx}_g$ and the \ac{Rx}~\cite[Eq.~(19)]{Jakumeit2026}, i.e., 
\begin{equation}\label{eqn:CIR_between_Tx_and_RX}
    h_{\mathrm{Tx}_g,w}(z_\mathrm{Rx},t;z_{\mathrm{Tx}_g})= \hspace*{-5mm}\sum_{\{P_k\in \mathcal{P}(\mathcal{S}_\mathrm{Tx}(\mathrm{Tx}_g),\mathcal{D}(p_w))\vert p_q,p_w\in P_k\}} \hspace*{-10mm}\gamma_{P_k} \bar{j}_k(z_\mathrm{Rx},t;z_{\mathrm{Tx}_g}),
\end{equation}
where the fraction of molecules $\gamma_{P_k}$ propagating through path $P_k$ is given as~\cite[Eq.~(22)]{Mosayebi2019}  
\begin{equation}\label{eqn:PathWeight}
    \gamma_{P_k} = \prod\limits_{\substack{p_i,b_m\in P_k,\\ p_i\in\mathcal{O}(b_m )}} \dfrac{Q_{i}}{\sum_{p_v\in\mathcal{O}(b_m)} Q_{v}}\,.
\end{equation}

\subsection{Receiver Model}
In this work, we focus on localization based on the signal received at a single sensor, i.e., \ac{Rx}. We model the sensor as a transparent molecule counting \ac{Rx} positioned in pipe $p_w$ and characterized by its length $l_{\mathrm{Rx}}\in(0,l_w]$ and its longitudinal center position $z_{\mathrm{Rx}}\in[0+l_{\mathrm{Rx}}/2,l_w-l_{\mathrm{Rx}}/2]$. 
Then, assuming the advective flux in pipe $p_w$ dominates the diffusive flux, the expected number of molecules at the \ac{Rx} due to a release of $M_g$ molecules at $\mathrm{Tx}_g$ follows as~\cite[Eqs.~(23), (24)]{Jakumeit2026}
\begin{align}\label{eqn:Nobs}
    N_{\mathrm{Rx},\mathrm{Tx}_g}(t)&=\frac{M_g}{\bar{u}_w}\int_{z_{\mathrm{Rx}}-l_{\mathrm{Rx}}/2}^{z_{\mathrm{Rx}}+l_{\mathrm{Rx}}/2}
    h_{\mathrm{Tx}_g,w}(z_w,t;z_{\mathrm{Tx}_g})\,\mathrm{d}z_w\nonumber \\
    &\approx\frac{M_g l_\mathrm{Rx}}{\bar{u}_w}h_{\mathrm{Tx}_g,w}(z_\mathrm{Rx},t;z_{\mathrm{Tx}_g}),
\end{align}
where the latter approximation is valid under the \ac{UCA}.

In practice, the \ac{Rx} does not have access to time-continuous molecule counts $N_{\mathrm{Rx},\mathrm{Tx}_g}(t)$. 
Instead it obtains every $T_\mathrm{s}$ seconds a noisy sample of $N_{\mathrm{Rx},\mathrm{Tx}_g}(t)$, which is inherently band-limited. 
In this paper, we neglect channel-induced counting noise in \eqref{eqn:sensor_output}, as $N_{\mathrm{Rx},\mathrm{Tx}_g}(t)$ will be typically very large and thus counting noise will be negligible~\cite{Jakumeit2025b}. 
However, we assume that the $k$-th sample is analyzed by a sensor which is subject to noise $N_\mathrm{s}[k]$, where $N_\mathrm{s}[k]$ is modeled as zero-mean \ac{AWGN} with variance $\sigma_\mathrm{s}^2$. 
Considering this model, where the sensor noise power is independent of the sampling frequency $f_{\mathrm{s}} = 1/T_{\mathrm{s}}$, we arrive at the following expression for the sensor response
\begin{equation}\label{eqn:sensor_output}
r[k] = N_{\mathrm{Rx},\mathrm{Tx}_g}(k T_\mathrm{s}) + N_\mathrm{s}[k] \defeq N_{\mathrm{Rx},\mathrm{Tx}_g}[k] + N_\mathrm{s}[k].
\end{equation}

%% file: sections/localization_algorithms_new.tex
We consider source localization in \acp{PN} with multiple \acp{Tx} and a single \ac{Rx}, where exactly one \textit{active} \ac{Tx}, $\mathrm{Tx}_g$, emits molecules in a single impulsive release\footnote{The latter assumption is justified in cases where emissions at the active \ac{Tx} are rare.}, see Fig.~\ref{fig:system_model}b). While the \ac{PN} topology and all \ac{Tx} positions are known, the active $\mathrm{Tx}_g$ is \textit{unknown}, and localization is defined as identifying it based on the received signal in~\eqref{eqn:sensor_output}.

\subsection{Matched Filter Bank Algorithm}\label{sec:algorithm}
We propose an \ac{MF}-based algorithm that aims to distinguish between individual \acp{Tx} based on the received signal by exploiting the known \ac{CIR} in~\eqref{eqn:CIR_between_Tx_and_RX} between any \ac{Tx} and the \ac{Rx}, see Fig.~\ref{fig:system_model}c).

\subsubsection{Signal Pre-Processing}\label{ssec:pre-processing}

For a single active $\mathrm{Tx}_g$ releasing $M_g$ molecules, the discrete expected number of molecules arriving at the \ac{Rx} in sampling interval $k$ is given as $N_{\mathrm{Rx},\mathrm{Tx}_g}[k]$, see~\eqref{eqn:Nobs}.
Since $M_g$ is an \ac{RV} whose realization is unknown at the \ac{Rx} and may vary over several orders of magnitude~\cite{foladori:SARS-CoV-2_from_faeces_to_wastewater_treatment_What_do_we_know_A_review, rose:The_Characterization_of_Feces_and_Urine_A_Review_of_the_Literature_to_Inform_Advanced_Treatment_Technology}, the amplitude of the received signal in~\eqref{eqn:sensor_output} cannot be reliably exploited for localization, see Figs.~\ref{fig:sewage_network}c) and \ref{fig:sewage_network}d). 
Normalizing the received signal by its maximum value helps to mitigate this problem, but due to the \ac{AWGN} at the \ac{Rx}, it is not possible to know the exact signal maximum. 
Therefore, the received signal is first filtered with a \ac{LP} filter $h_\mathrm{LP}[k]$, whose cut-off frequency is chosen based on the frequency contents of the known \acp{CIR}, resulting in the filtered received signal\vspace*{-1mm}
\begin{equation}\label{eqn:lowpass_filtered_received_signal}
r_{\mathrm{LP}}[k]\hspace*{-.5mm}=\hspace*{-.5mm} r[k] \hspace*{-.5mm}*\hspace*{-.5mm} h_{\mathrm{LP}}[k]\hspace*{-.5mm}\overset{\eqref{eqn:sensor_output}}{=}\hspace*{-.5mm} \underbrace{N_{\mathrm{Rx},\mathrm{Tx}_g}[k]\hspace*{-.5mm}*\hspace*{-.5mm}h_{\mathrm{LP}}[k]}_{\defeq \,N_{\mathrm{Rx},\mathrm{Tx}_g,\mathrm{LP}}[k]} \hspace*{-.5mm}+\hspace*{-.5mm} \underbrace{N_{\mathrm{s}}[k]\hspace*{-.5mm}*\hspace*{-.5mm}h_{\mathrm{LP}}[k]}_{\defeq \,N_{\mathrm{s},\mathrm{LP}}[k]},\vspace*{-1mm}
\end{equation}
where $*$ denotes convolution with respect to time, and $N_{\mathrm{Rx},\mathrm{Tx}_g,\mathrm{LP}}[k]$ and $N_{\mathrm{s},\mathrm{LP}}[k]$ are the \ac{LP}-filtered signal and sensor noise components of $r_\mathrm{LP}[k]$, respectively.
Afterwards, $r_{\mathrm{LP}}[k]$ is normalized by it's maximum value, which is now much less impacted by noise, yielding an \ac{LP}-filtered, normalized version of the received signal
\begin{equation}\label{eqn:normalized_filtered_receiver_input}
\tilde{r}_{\mathrm{LP}}[k] = \frac{r_\mathrm{LP}[k]}{\max\limits_k(r_\mathrm{LP}[k])}\,.
\end{equation}
The normalization of the received signal removes all information about the number of emitted molecules in ${r}_\mathrm{LP}[k]$.

\subsubsection{Matched Filter Design}\label{ssec: MF algorithm}

The signal used to construct the \ac{MF} for a given $\mathrm{Tx}_g$ does not rely directly on the noisy received signal in~\eqref{eqn:sensor_output}. Rather, it is calculated using the analytical channel model in \eqref{eqn:Nobs}. 
Since the \ac{MF} is applied \textit{after} the \ac{LP} filtering of the received signal, in the first step of the \ac{MF} design, the number of observed molecules in~\eqref{eqn:Nobs} is \ac{LP}-filtered with the same filter $h_\mathrm{LP}[k]$ as in~\eqref{eqn:lowpass_filtered_received_signal}. 
This yields the \ac{LP}-filtered expected number of molecules $N_{\mathrm{Rx},\mathrm{Tx}_g,\mathrm{LP}}[k]$ arriving  at the \ac{Rx} (see also \eqref{eqn:lowpass_filtered_received_signal}). To account for the normalization of the received signal in~\eqref{eqn:normalized_filtered_receiver_input}, we also normalize the expected number of molecules by its maximum value, i.e.,
\begin{equation}\label{eqn:Normalized signal}
\tilde{N}_{\mathrm{Rx},\mathrm{Tx}_g,\mathrm{LP}}[k] = \frac{N_{\mathrm{Rx},\mathrm{Tx}_g,\mathrm{LP}}[k]}{\max\limits_k (N_{\mathrm{Rx},\mathrm{Tx}_g,\mathrm{LP}}[k])}\,.
\end{equation}

Because we use an \ac{LP} filter in~\eqref{eqn:lowpass_filtered_received_signal}, the originally white sensor noise becomes colored.
To ensure the \ac{MF} accounts for this, we first compute the colored noise \ac{ACF} as
\begin{equation}\label{eqn:Noise acf}
R_{nn}[k] =  \sigma_\mathrm{s}^2\delta[k]*\, h_{\mathrm{LP}}[k]*h_{\mathrm{LP}}[-k]=\sigma_\mathrm{s}^2\, h_{\mathrm{LP}}[k]*h_{\mathrm{LP}}[-k],
\end{equation}
where $\delta[k]$ denotes the Kronecker delta function.
From the \ac{ACF} in \eqref{eqn:Noise acf}, the colored noise autocorrelation matrix follows as a Toeplitz matrix~\cite{Hayes1996}
\begin{equation}
\setlength{\arraycolsep}{1.5pt}
 \mathbf{R}_{nn} =
    \begin{bmatrix}
    R_{nn}[0] & R_{nn}[1] & R_{nn}[2] & \cdots & R_{nn}[M\hspace*{-.5mm}-\hspace*{-.5mm}1] \\
    R_{nn}[1] & R_{nn}[0]  & R_{nn}[1] & \cdots & R_{nn}[M\hspace*{-.5mm}-\hspace*{-.5mm}2] \\
    \vdots    & \vdots     & \vdots     & \ddots & \vdots \\
    R_{nn}[M\hspace*{-.5mm}-\hspace*{-.5mm}1] & R_{nn}[M\hspace*{-.5mm}-\hspace*{-.5mm}2] & R_{nn}[M\hspace*{-.5mm}-\hspace*{-.5mm}3] & \cdots & R_{nn}[0]
    \end{bmatrix}\,.
\end{equation}
Next, we represent the expected, \ac{LP}-filtered, and normalized signal $\tilde{N}_{\mathrm{Rx},\mathrm{Tx}_g,\mathrm{LP}}[k]$ as a vector of finite length who's elements are the sampled signal values, i.e.,
\begin{align}
    \mathbf{\tilde{N}}_{\mathrm{Rx},\mathrm{Tx}_g,\mathrm{LP}} = [\tilde{N}_{\mathrm{Rx},\mathrm{Tx}_g,\mathrm{LP}}[0],\ldots,\tilde{N}_{\mathrm{Rx},\mathrm{Tx}_g,\mathrm{LP}}[M-1]]^\top,
\end{align}
where $(\cdot)^\top$ denotes the transpose operator and $M$ is chosen large enough such that the expected signal has decayed to zero.

\begin{figure}
    \centering
    \includegraphics[width=\linewidth]{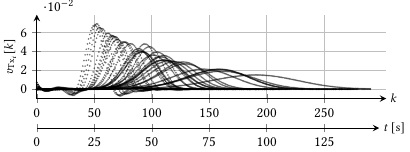}
    \caption{\acp{MF} $v_{\mathrm{Tx}_i}[k]$ used for source localization in the \ac{PN} shown in Fig.~\ref{fig:sewage_network}a) with $U=33$ \acp{Tx} at a sensor sampling frequency of $f_\mathrm{s}=2\,\mathrm{Hz}$.}
    \Description[]{}
    \label{fig:matched_filters}
\end{figure}

Finally, the \ac{MF} for $\mathrm{Tx}_g$ is\footnote{The \ac{MF} design in~\eqref{eqn:MF} is motivated by radar signal processing~\cite{Richards2005}, where such filters are known as \textit{Capon} or \textit{Minimum Variance Distortionless Response (MVDR)} beamformers. The \ac{Tx} decision rule in~\eqref{eqn:Tx_decision} is motivated by the same framework.} a vector of length $M$~\cite[Eq.~(9.19)]{Richards2005}
\begin{equation}\label{eqn:MF}
\mathbf{v}_{\mathrm{Tx}_g}
=
\frac{
\mathbf{R}_{nn}^{-1}\,\mathbf{\tilde{N}}_{\mathrm{Rx},\mathrm{Tx}_g,\mathrm{LP}}
}{
\mathbf{\tilde{N}}_{\mathrm{Rx},\mathrm{Tx}_g,\mathrm{LP}}^\top\,
\mathbf{R}_{nn}^{-1}\,
\mathbf{\tilde{N}}_{\mathrm{Rx},\mathrm{Tx}_g,\mathrm{LP}}
}, \quad g \in \{1,\ldots , U\},
\end{equation}
where the numerator maximizes the \ac{SNR} for the received signal from $\mathrm{Tx}_g$, and the denominator is a normalization scalar
ensuring a maximum \ac{MF} output amplitude of 1 when convolved with the corresponding received signal.
We denote the $k$-th element of the \ac{MF} as $v_{\mathrm{Tx}_g}[k]$, $k\in\{0,\ldots,M-1\}$. See Fig.~\ref{fig:matched_filters} for exemplary \acp{MF}. Negative values arise from the \ac{LP} filtering in~\eqref{eqn:lowpass_filtered_received_signal}.

\subsubsection{Matched Filtering of Received Signal}
To perform the matched filtering, we collect for all time steps in the observation interval the past $M$ samples of the \ac{LP}-filtered, normalized received signal $\tilde{r}_\mathrm{LP}[k]$ in a vector $\mathbf{\tilde{r}}_{\mathrm{LP}}[k]$ according to
\begin{equation}
    \mathbf{\tilde{r}}_{\mathrm{LP}}[k] = [\tilde{r}_{\mathrm{LP}}[k],\ldots,\tilde{r}_{\mathrm{LP}}[k-M+1]]^\top.\label{eqn:received_vector}
\end{equation}
To identify the active \ac{Tx}, the signal in~\eqref{eqn:received_vector} is filtered by a filter bank comprising the \acp{MF} of all \acp{Tx} for all time steps in the observation interval, i.e.,
\begin{align}\label{eqn:matched_filter}
y_i[k]= \mathbf{v}_{\mathrm{Tx}_i}^\top\,\mathbf{\tilde{r}}_\mathrm{LP}[k],\quad i\in\{1,\ldots ,U\}\,.
\end{align}
Then, the \ac{Tx} whose MF output has the maximum amplitude closest to 1, denoted by $\mathrm{Tx}_{i^*}$, is selected as the candidate active \ac{Tx}, i.e.,
\begin{align}\label{eqn:Tx_decision}
i^* \hspace*{-.6mm}=\hspace*{-.5mm} \arg\min_i \left| y^\mathrm{max}_i \hspace*{-.6mm}-\hspace*{-.5mm} 1 \right|,\hspace*{2mm}{y^\mathrm{max}_i}\hspace*{-.5mm}=\hspace*{-.5mm}\max_{k}(y_i[k]),\hspace*{2mm} i \in \{1,\dots,U\}.
\end{align}

Lastly, to prevent pure noise at the \ac{Rx} from being misidentified as a signal, the ratio of the peak power to noise variance, $(y_{i^*}^\mathrm{max})^2/\hat{\sigma}^2$ is computed for the \ac{MF} output corresponding to the candidate active \ac{Tx}.
The noise variance $\hat{\sigma}^2$ is estimated within an observation window
around the peak, where samples are included only once their amplitude drops
below $10\,\%$ of the peak amplitude.

\subsubsection{Basic Localization Algorithm}\label{sssec:algorithm}

The individual steps of the localization algorithm are illustrated in Fig.~\ref{fig:system_model}c). The received signal is first passed through an \ac{LP} filter, see~\eqref{eqn:lowpass_filtered_received_signal}, and then normalized by its maximum value, see~\eqref{eqn:normalized_filtered_receiver_input}. 
Subsequently, an \ac{MF} bank comprising the \acp{MF} of all \acp{Tx}, see~\eqref{eqn:matched_filter}, is applied. Finally, the \ac{MF} whose maximum output is closest to~1 is identified, see~\eqref{eqn:Tx_decision}, and the \ac{Tx} corresponding to this filter is declared as the localized source only if the peak power to noise variance ratio surpasses a threshold. For our simulations, we have set this threshold to be equal to 20.

\subsection{Clustering Likely Confused Sources}\label{sec:clustering}

Localization based on an \ac{MF} bank performs well when the received signals associated with different \acp{Tx} exhibit sufficiently distinct temporal \textit{shapes}. 
However, depending on the \ac{PN} topology and the spatial proximity of the \acp{Tx}, signals from different \acp{Tx} may have highly similar shapes, see Section~\ref{sec:Results}. 
In such cases, it can be advantageous to group \acp{Tx} with similar \acp{CIR} into clusters and perform localization at a cluster-level, where localization is considered successful if the estimated \ac{Tx} belongs to the same cluster as the active \ac{Tx}. 
Although this approach may not identify the exact active \ac{Tx}, it narrows the set of prospective \acp{Tx} to a small subset of all \acp{Tx}.

In this work, clustering is performed using the Louvain community detection algorithm~\cite{Blondel2008} applied to a graph whose nodes represent all \acp{Tx} in the \ac{PN} and whose edge weights quantify the similarity between their corresponding \acp{CIR}. This similarity graph can be constructed either empirically from the \ac{CM} or analytically from the \ac{CSM}.
Because the Louvain algorithm chooses clusters that maximize \textit{modularity}, a measure for how well a graph is separated into tightly-knit groups, it groups very similar \acp{CIR} together as these have high edge weights. 
In the following, we describe the construction of the \ac{CM} and \ac{CSM}, and the subsequent evaluation of the clustering.

\subsubsection{Confusion Matrix}\label{ssec:confusion_matrix}
The \ac{CM} is denoted by $\mathbf{CM}\in[0,1]^{U\times U}$. It is constructed by applying the localization algorithm in Section~\ref{sssec:algorithm} $N_\mathrm{sim}$ times for each $\mathrm{Tx}_i\in\mathcal{N}_\mathrm{Tx}$, using independent noisy realizations of the received signal in~\eqref{eqn:sensor_output}. 
After a total of $U\cdot N_\mathrm{sim}$ runs, the entry $CM_{ij}\in[0,1]$ of $\mathbf{CM}$ represents the fraction of simulation runs where, instead of the active $\mathrm{Tx}_i$, $\mathrm{Tx}_j$ was classified as the signal source. Entries on the main diagonal represent the fractions of simulation runs where the active \ac{Tx} was correctly identified.

We find that binarizing $\mathbf{CM}$ reduces variability across
simulation runs and improves clustering quality
(see Section~\ref{sssec:clustering_quality}).
The binarized matrix $\mathbf{CM}^\mathrm{b}$ is obtained by setting
entries with fewer than $5\,\%$ confusions to zero and all others to
one, i.e.,
\begin{equation}
    CM_{ij}^\mathrm{b} = \begin{cases}
    1, &\text{if}\quad CM_{ij} \geq 0.05 \\
    0, &\text{if}\quad CM_{ij} < 0.05
    \end{cases}\,.
\end{equation}
Matrix $\mathbf{CM}^\mathrm{b}$ is interpreted as a directed graph,
on which the Louvain community detection algorithm~\cite{Blondel2008} is applied
to obtain clusters.

\subsubsection{Cosine Similarity Matrix}
Contrary to the \ac{CM}, the \ac{CSM}, denoted by $\mathbf{CSM}\in[0,1]^{U\times U}$, does not rely on empirical simulations. 
It is given by the pairwise cosine similarities of the \ac{Tx} \acp{CIR}, i.e.,
\begin{equation}\label{eqn:cosine_similarity_matrix}
CSM_{ij} =
\frac{\int_{-\infty}^{\infty} 
h_{\mathrm{Tx}_i,w}(z_\mathrm{Rx},t;z_{\mathrm{Tx}_i}) \, 
h_{\mathrm{Tx}_j,w}(z_\mathrm{Rx},t;z_{\mathrm{Tx}_j}) \, \mathrm{d}t
}
{
\sqrt{\int_{-\infty}^{\infty} h_{\mathrm{Tx}_i,w}^2(z_\mathrm{Rx},t;z_{\mathrm{Tx}_i}) \, \mathrm{d}t} \;
\sqrt{\int_{-\infty}^{\infty} h_{\mathrm{Tx}_j,w}^2(z_\mathrm{Rx},t;z_{\mathrm{Tx}_j}) \, \mathrm{d}t}
}\,.
\end{equation}
Binarization is performed similarly as for the \ac{CM}, with an adjusted threshold, as the cosine similarity metric assigns values close to one for similar signals, i.e.,
\begin{equation}\label{eqn:csm_bin}
    CSM_{ij}^\mathrm{b} = \begin{cases}
    1, &\text{if}\quad CSM_{ij} \geq{0.95} \\
    0, &\text{if}\quad CSM_{ij} <{0.95}
    \end{cases}\,.
\end{equation}
Note that $\mathbf{CSM}^\mathrm{b}$ is symmetric and is interpreted as an undirected graph in the Louvain community algorithm~\cite{Blondel2008}, which is applied to obtain clusters of likely confused \acp{Tx}.

\subsubsection{Clustering Quality}\label{sssec:clustering_quality}

\acp{Tx} are considered well clustered if \acp{CIR} within each cluster are highly similar, while differing significantly across different clusters. To quantify clustering quality, we employ the silhouette score~\cite{Peter1987} based on the squared Euclidean distance, denoted by $\left\|\cdot \right\|_2$, between the normalized \acp{CIR}, i.e.,
\begin{align}
&d(\mathrm{Tx}_i,\mathrm{Tx}_j)
=
\left\|
\tilde{h}_{\mathrm{Tx}_i,w}(z_{\mathrm{Rx}},t;z_{\mathrm{Tx}_i})
-
\tilde{h}_{\mathrm{Tx}_j,w}(z_{\mathrm{Rx}},t;z_{\mathrm{Tx}_j})
\right\|_{2}^2,\label{eqn:impulse_response_normalized} \\
    &\text{with}\quad\tilde{h}_{\mathrm{Tx}_i,w}(z_\mathrm{Rx},t;z_{\mathrm{Tx}_i})=\frac {h_{\mathrm{Tx}_i,w}(z_\mathrm{Rx},t;z_{\mathrm{Tx}_i})}{\max\limits_{t}({h_{\mathrm{Tx}_i,w}(z_\mathrm{Rx},t;z_{\mathrm{Tx}_i}))}}.
\end{align}
The silhouette score ranges from $-1$ to $1$, with higher values indicating compact, well-separated clusters and lower values reflecting poor clustering quality. To compare the similarity of the clustering results obtained from $\mathbf{CM}^\mathrm{b}$ and $\mathbf{CSM}^\mathrm{b}$, we use the \ac{ARI}~\cite[Eq.~(21)]{Hubert1985}, ranging from $-0.5$ to $1$, where a value of $1$ indicates identical clustering based on $\mathbf{CM}^\mathrm{b}$ and $\mathbf{CSM}^\mathrm{b}$.

%% file: sections/results.tex
To investigate the performance of the proposed localization approaches, we first introduce a real-world \ac{SN} as an exemplary application domain and then assess the localization performance. 

\begin{figure*}
    \centering
    \includegraphics[width=\linewidth]{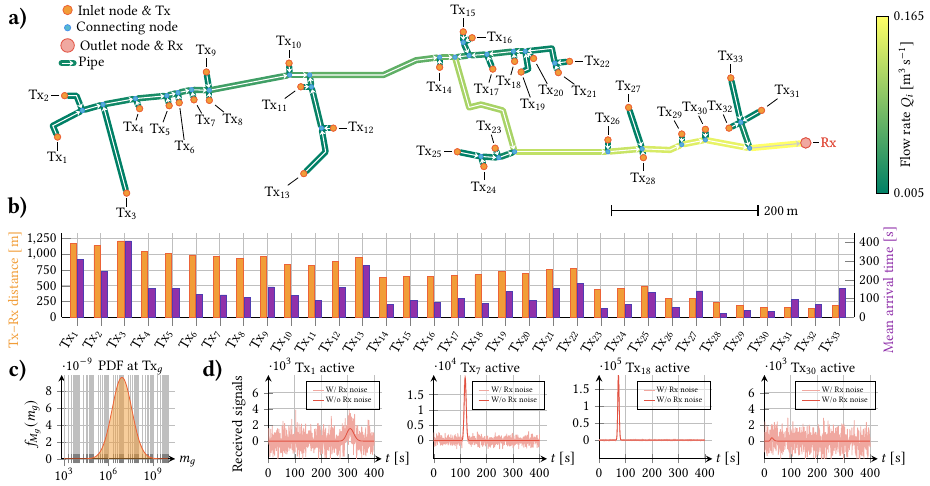}\vspace*{-3mm}
    \caption{a) Topology of a \ac{SN} in Zolkiewka Commune, Poland~\cite{Nawrot2018}. Pipe lengths $l_i$ are drawn to scale (see scale bar) and $r_i=\SI{5.5}{\centi\meter},\forall p_i$. Inlet, connecting, and outlet node positions are illustrated in orange, blue, and red, respectively. At each inlet node $n_{\mathrm{in},g}\in\mathcal{N}_\mathrm{in}$, a $\mathrm{Tx}_g$ is present with $Q_{\mathrm{in},g}=\SI{5e-3}{\meter\cubed\per\second}$. Flow directions are given by arrows, flow rates are color-coded in the edge colors. The \ac{Rx} is located at the outlet node. b)~\ac{Tx}-\ac{Rx} distances (orange) and \ac{Tx}-\ac{Rx} path mean arrival times (purple) for all \acp{Tx}. c)~Log-normal \ac{PDF} for the number of released viral particles $M_g$ at the active $\mathrm{Tx}_g$. d)~Four exemplary received signals from different active \acp{Tx} with and without additive \ac{Rx} noise.}
    \Description[]{}
    \label{fig:sewage_network}
\end{figure*}

\subsection{Sewage Pipe Network}

We simulate part of a \ac{SN} in the Zolkiewka Commune, Poland~\cite{Nawrot2018}, as shown in Fig.~\ref{fig:sewage_network}a). 
We make the following simplifying assumptions: 
Each house connected to the sewage system is modeled as an inlet node $n_{\mathrm{in},g}$ with a constant inflow rate $Q_{\mathrm{in},g}=\SI{5e-3}{\meter\cubed\per\second}$, resulting in realistic flow rates within the \ac{PN}, see Fig.~\ref{fig:sewage_network}a). 
At each inlet $n_{\mathrm{in},g}$, a potentially active $\mathrm{Tx}_g$ is placed at $z_{\mathrm{Tx}_g}=0$, representing a source of viral markers. 
Molecule transport in the \ac{SN} is assumed to be advection-diffusion-driven with diffusion coefficient {$D=\SI{0.2}{\meter\squared\per\second}$}~\cite{Sonnenwald2023}, while gravitational effects and half-filled pipes typical of \acp{SN} are neglected~\cite{Nawrot2018}.

At the active $\mathrm{Tx}_g$, we assume a single impulsive release of viral markers.
The number of released markers, denoted by $M_g$, is modeled as log-normally
distributed, i.e., $\mathrm{ln}(M_g)\sim\mathcal{N}(\mu_\mathrm{Tx},\sigma_\mathrm{Tx}^2)$,
with mean $\mu_\mathrm{Tx}=18.69$ and variance $\sigma_\mathrm{Tx}^2 = 2.46$
\cite{foladori:SARS-CoV-2_from_faeces_to_wastewater_treatment_What_do_we_know_A_review,
rose:The_Characterization_of_Feces_and_Urine_A_Review_of_the_Literature_to_Inform_Advanced_Treatment_Technology},
see Fig.~\ref{fig:sewage_network}c).

At the \ac{Rx}, a default sampling rate of $f_{\mathrm{s}}=2\,\mathrm{Hz}$ and noise power
$\sigma_\mathrm{s}^2=10^3$ are assumed.
We emphasize that \ac{MIGHT} serves as a simplified proxy for sewage
transport, as our focus is on generic localization techniques in \acp{PN}
rather than high-fidelity \ac{SN} modeling.

\subsection{Simulation Results}
Below, we first present results on the \ac{CM} and \ac{CSM} and investigate the \ac{Tx} clustering. Subsequently, we illustrate the localization accuracy for various sensor noise powers and sampling frequencies.

\subsubsection{Transmitter Confusion Matrix and Cosine Similarity Matrix}

\begin{figure}
    \centering
    \includegraphics[width=\linewidth]{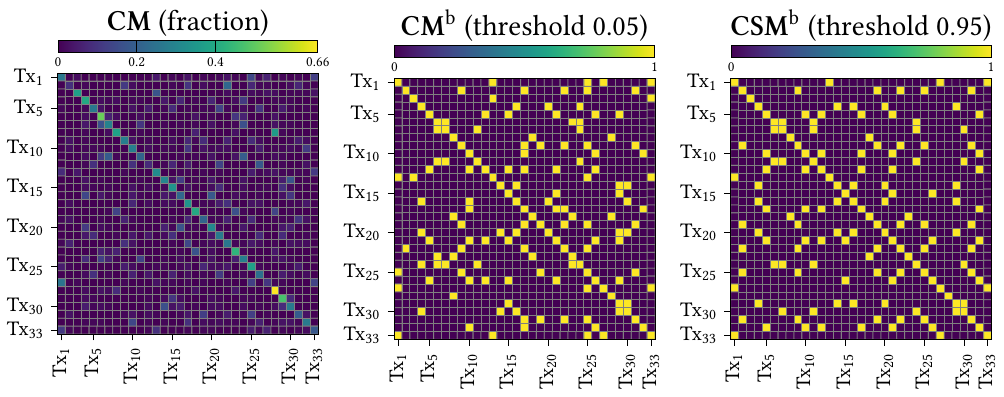}\vspace*{-3mm}
    \caption{Likely confused \acp{Tx}, illustrated using the empirically determined \ac{CM} (left) and its binarized version (middle). Confusions predicted by the binarized \ac{CSM} are shown on the right.}
    \Description[]{}
    \label{fig:all_matrices}
\end{figure}

Due to the similarity of the temporal shapes of the received signals associated with different \acp{Tx}, see Fig.~\ref{fig:clustering_and_molecule_types}, several \ac{Tx} pairs are likely confused by the localization algorithm in Section~\ref{sssec:algorithm}. For $N_\mathrm{sim}=100$ simulation runs per \ac{Tx}, the resulting confusion rates are shown in Fig.~\ref{fig:all_matrices} using the \ac{CM} (left) and its binarized version (middle). In the \ac{CM}, the observed confusion rates off the main diagonal range from $0\,\%$ to $36\,\%$, with particularly high confusions between $\mathrm{Tx}_{8}$ and $\mathrm{Tx}_{28}$ ($36\,\%$). Note that $\mathbf{CM}^\mathrm{b}$ is approximately symmetric.

The right plot in Fig.~\ref{fig:all_matrices} shows the confusions predicted by $\mathbf{CSM}^\mathrm{b}$.
The agreement between clustering based on $\mathbf{CM}^\mathrm{b}$ and two other clustering methods is quantified in Table~\ref{tab:Clustering} using the \ac{ARI}, while clustering quality is evaluated via the silhouette score in~\eqref{eqn:impulse_response_normalized}.
Both clustering methods based on $\mathbf{CM}^\mathrm{b}$ and $\mathbf{CSM}^\mathrm{b}$ yield good silhouette scores ($0.49$ and $0.60$, respectively).
Moreover, $\mathbf{CM}^\mathrm{b}$ and $\mathbf{CSM}^\mathrm{b}$ show good
agreement ($\mathrm{ARI}=0.70$), confirming that the \ac{CSM} is a useful
proxy for the actual \ac{CM}.
As a baseline, we include K-means clustering based on the path means $\bar{\mu}_g$, with the number of clusters set equal to that obtained with the $\mathbf{CM}^\mathrm{b}$ and $\mathbf{CSM}^\mathrm{b}$ methods.
Compared to these approaches, K-means yields poorer clusters (silhouette score $-0.2763$) and does not agree with the \ac{CM} ($\mathrm{ARI}=-0.0053$).

\begin{table}
\caption{Clustering evaluation metrics.\vspace{-4mm}}
\label{tab:Clustering}
\begin{tabular}{l|c|c}
\bottomrule
Clustering  & Silhouette score & ARI \\
\toprule
    $\mathbf{CM}^\mathrm{b}$ & \ $0.49$  & \ $1.00$\\
    $\mathbf{CSM}^\mathrm{b}$ & \ $0.60$  & \ $0.70$\\
    K-means ($\bar{\mu}_g$) & $-0.2763$  & $-0.0053$\\
\bottomrule
\end{tabular}
\end{table}

\subsubsection{Spatial Spread of Source Clusters}\label{sec:molecule_assignment}

\begin{figure}
    \centering
    \includegraphics[width=\linewidth]{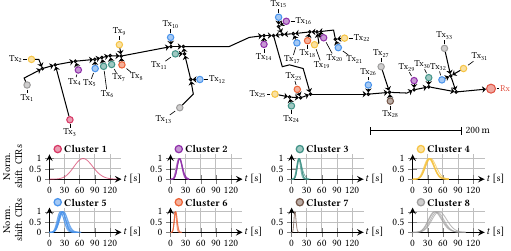}\vspace*{-3mm}
    \caption{Spatial distribution of \ac{Tx} clusters. Below the \ac{SN}, the normalized and time-shifted \acp{CIR} of all clusters are shown.}
    \Description[]{}
    \label{fig:clustering_and_molecule_types}
\end{figure}

In the top part of Fig.~\ref{fig:clustering_and_molecule_types}, the cluster assignment of each \ac{Tx} is indicated by its color, as determined using $\mathbf{CSM}^\mathrm{b}$ in \eqref{eqn:csm_bin}.
The bottom part shows the normalized and time-shifted \acp{CIR} of all \acp{Tx} within each cluster.
The time shift accounts for the unknown emission time at the \ac{Tx}. 
Interestingly, \textit{while the \acp{CIR} of \acp{Tx} within a cluster are very similar, the corresponding \acp{Tx} are often not geographically close}.
This is because many parts of the \ac{SN} exhibit heterogeneous flow rates in their immediate neighborhood, see Fig.~\ref{fig:sewage_network}a), leading to strongly varying \acp{CIR} of the \acp{Tx} in these neighborhoods. This can also be observed in Fig.~\ref{fig:sewage_network}b), where it becomes clear that the \ac{Tx}-\ac{Rx} distances are not necessarily correlated to mean arrival times, due to varying flow rates.
This has direct implications when using an \ac{MF}-based localization algorithm to identify, e.g., an infected individual in a community. Investigating only the members of the identified household and geographically neighboring households is insufficient. Instead, households that are \textit{close} in terms of molecule propagation time to the \ac{Rx} must be considered, as the molecule release time is not known.

\subsubsection{Localization Accuracy}

\begin{figure}
    \centering
    \includegraphics[width=\linewidth]{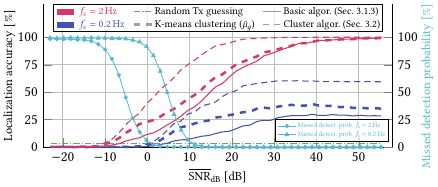}\vspace*{-3mm}
    \caption{Localization accuracy of the proposed algorithms and missed detection probability  at different sensor sampling rates.}
    \Description[]{}
    \label{fig:localization_accuracy}
\end{figure}

Firstly, in Fig.~\ref{fig:localization_accuracy}, we show the localization accuracy of the \ac{MF}-based localization algorithm proposed in Section~\ref{sssec:algorithm} as a function of the \ac{SNR} for both a sampling rate of $f_\mathrm{s} = 2\,\mathrm{Hz}$ (pink solid line) and $f_\mathrm{s} = 0.2\,\mathrm{Hz}$ (blue solid line).
For this algorithm, the localization accuracy is defined as the percentage of simulation runs in which the active \ac{Tx} is correctly identified.
Secondly, Fig.~\ref{fig:localization_accuracy} shows the localization accuracy for the \ac{CSM}-based clustering approach in Section~\ref{sec:clustering} for both sampling rates (thin dashed pink and blue lines) and clustering baselines based on K-means clustering (thick dashed pink and blue lines).
In this setting, the localization accuracy is defined as the percentage of simulation runs in which the identified \ac{Tx} lies within the cluster containing the active \ac{Tx}.
Thirdly, Fig.~\ref{fig:localization_accuracy} shows the missed detection probability computed from the simulation runs (cyan solid lines), i.e., the probability that a signal is confused with noise and no localization is performed, which only depends on the sampling rate and the \ac{SNR}.

The \ac{SNR} is obtained by first calculating, for each \ac{Tx}, the ratio between the expected received signal's~\eqref{eqn:Nobs} average power (using the mean number of released molecules $\bar{M}_g$) and the sensor noise power, and then averaging over all $U$ \acp{Tx} in the \ac{PN}, i.e.,
\vspace*{-1mm}
\begin{equation}
    \overline{\mathrm{SNR}}_\mathrm{dB} \hspace*{-.5mm}=\hspace*{-.5mm} 10\log_{10}\left( \frac{1}{U}\hspace*{-1mm}\sum_{\mathrm{Tx}_g\in\mathcal{N}_\mathrm{Tx}}\hspace*{-2mm}\frac{\frac{1}{t^2_g-t^1_g}\hspace*{-.5mm}\int_{t^1_g}^{t^2_g} \hspace*{-.5mm}N^2_{\mathrm{Rx,Tx}_g}(z_{\mathrm{Rx}},t;z_{\mathrm{Tx}_g})\mathrm{d}t}{\sigma_s^2}\right),
\end{equation}
\vspace*{-1mm}where $[t^1_g,t^2_g]$ represents the time interval during which the signal of $\mathrm{Tx}_g$ carries most of its energy. In our simulations, $\overline{\mathrm{SNR}}_\mathrm{dB}$ is varied by varying the sensor noise power $\sigma_\mathrm{s}^2$.

Starting with the sensor with sampling frequency $f_\mathrm{s}=2\,\mathrm{Hz}$ (pink lines), the basic algorithm outperforms random guessing (green line, accuracy $1/U=1/33$) already at $\overline{\mathrm{SNR}}_\mathrm{dB}=-10\,\mathrm{dB}$. 
For lower \acp{SNR}, the algorithm cannot distinguish signal from noise and thus fails to detect any \acp{Tx}, resulting in large missed detection probabilities (cyan line with round markers) and zero correct identifications.
As the \ac{SNR} increases, the algorithm eventually achieves perfect localization accuracy.
When considering the ability to predict the correct cluster (thin pink dashed line), we observe a gain of more than $10\,\mathrm{dB}$ compared to predicting the exact \ac{Tx} (pink solid line), confirming the intuition from Section~\ref{sec:clustering} that most confusions occur in small clusters of \acp{Tx}.
Moreover, at both sampling rates, the clustering algorithm from Section~\ref{sec:clustering} (thin dashed lines) outperforms the algorithm based on K-means clustering (thick dashed lines).
At the lower sampling rate of $f_\mathrm{s}=0.2\,\mathrm{Hz}$ (blue lines), both the basic and clustering algorithms achieve significantly lower accuracy than for $f_\mathrm{s}=2\,\mathrm{Hz}$.
This is due to two effects: the missed detection probability shifts by $\sim10\,\mathrm{dB}$ to the right (see cyan lines), and even when it
reaches zero, perfect localization accuracy cannot be achieved
independent of the \ac{SNR}, since only limited information about the signal shape
can be recovered at lower sampling rates.

%% file: sections/conclusion.tex
In this work, we proposed the first analytical framework for molecule source
localization in advection-diffusion-driven \acp{PN} with known topology based on the \ac{MIGHT} channel model.
The proposed \ac{MF}-based approach reliably identifies a single active \ac{Tx} among
multiple known \acp{Tx} from the signal of a single \ac{Rx}, as illustrated
for a model of a real-world \ac{SN}.
Furthermore, frequently confused \acp{Tx} can be grouped into clusters that
remain reliably distinguishable, drastically reducing the number of \acp{Tx}
requiring manual inspection in, e.g., epidemiological applications.
Future work will study the transferability of the framework beyond
\acp{SN} across application domains and scales, and investigate
inter-\ac{Tx} interference and experimental validation in branched
\ac{MC} testbeds.